# Dot-Diffused Halftoning with Improved Homogeneity

Yun-Fu Liu, *Member*, IEEE and Jing-Ming Guo, *Senior Member*, IEEE

Department of Electrical Engineering,
National Taiwan University of Science and Technology,
Taipei, Taiwan
E-mail: yunfuliu@gmail.com, jmguo@seed.net.tw

**ABSTRACT**
Compared to the error diffusion, dot diffusion provides an additional pixel-level parallelism for digital halftoning. However, even though its periodic and blocking artifacts had been eased by previous works, it was still far from satisfactory in terms of the blue noise spectrum perspective. In this work, we strengthen the relationship among the pixel locations of the same processing order by an iterative halftoning method, and the results demonstrate a significant improvement. Moreover, a new approach of deriving the averaged power spectrum density (APSD) is proposed to avoid the regular sampling of the well-known Bartlett's procedure which inaccurately presents the halftone periodicity of certain halftoning techniques with parallelism. As a result, the proposed dot diffusion is substantially superior to the state-of-the-art parallel halftoning methods in terms of visual quality and artifact-free property, and competitive runtime to the theoretical fastest ordered dithering is offered simultaneously.

## 1. INTRODUCTION

Digital halftoning [1] is a technique for converting continuous-tone images into binary images. These binary images resemble the original images when viewed from a distance because of the low-pass nature of the human visual system (HVS). This technique has been utilized widely in rendering an image with limited colors to yield the perceptual illusion of more colors. So far, many commercial applications have been introduced in the market such as document printing and electronic paper (e-paper) displays. In general, the properties of halftones can be categorized into blue- or green-noise to render the frequency of dot appearance for various printers. For instance, inkjet printers exploit the advantage of blue-noise halftoning for a better illusion of a given shade of color [2]. Conversely, laser printers lean to consider green-noise halftoning, because of the unstable printed dots induced by the electrophotography printing process [3]. Another perspective of the classification considers their processing types: 1) Point process - ordered dithering [1], [4]-[5]; 2) neighborhood process - error diffusion [6]-[8], and dot diffusion [9]-[12]; 3) iterative process - direct binary search [13]-[14] and electrostatic halftoning [15]. Among these, iterative methods provide the best halftone texture, yet processing efficiency is their major issue for the complex updating process. In addition, methods involved neighborhood processing normally achieve the second best image quality in terms of the dot homogeneity and processing efficiency. This type of methods adaptively determines the dot distribution by considering the influence from the neighborhood as similar to that of the iterative methods, yet simply one-pass processing is required rather than the iterative strategy. In this category, as opposed to the error diffusion, dot diffusion further exploits the parallelism for a higher processing efficiency. Yet, the inherent neighborhood processing still significantly impedes the processing speed compared to that of the ordered dithering which simply requires point-by-point thresholding operation.

Specifically, dot diffusion which was first proposed in Knuth's work [16], reaping the benefits of parallelism through the use of the class matrix (CM) and diffused matrix (DM). Formerly, in Guo-Liu's work [10], a tone-similarity improvement strategy was proposed with a pair of co-optimized CM and DM for a higher image similarity. Yet, the periodic pattern still interferes the visual perception, and thus degrades the visual quality. To suppress the periodicity, Lippens and Philips [11] proposed the "grid diffusion" to enlarge the size of a CM for a greater spatial period of the duplicated textures, in which the grid was composed of a group of CMs. In their study, a grid of size 128×128 was constructed by 16×16 CMs of size 8×8. Subsequently, the near-aperiodic dot diffusion (NADD) [12] utilized a new class tiling (CT) designed dot diffusion to obtain aperiodic halftone patterns. The periodicity was further improved by manipulating the CT with rotation, transpose, and alternatively shifting operations with one pair of the optimized CM and DM. Yet, even the above existing methods have significantly suppressed the periodic artifacts. The corresponding halftone patterns still have unstable spectrum property, which ends up with an unstable tone rendering capability.

To further improve the visual quality based upon the prior arts, we found that the bottleneck leading to the above unstable tone rendering is caused by the use of the CM. We also found that it can be significantly improved by emphasizing the spatial relationship among the same processing orders in the CT. In this work, the CT is optimized with the dual-metric direct binary search (DMDBS) [13] for a great spectrum stability. Subsequently, to optimize the parameters of the proposed dot diffusion, the influences of the cost function selection and the use of Bartlett's procedure for spectrum error are discussed. As documented in the simulation results, the proposed method is substantially superior to the former dot diffusion methods in terms of visual quality and processing efficiency. Moreover, artifact-free property can be endorsed in contrast to the state-of-the-art ordered dithering methods. Meanwhile, in contrast to the DMDBS which is known for its excellent dot rendering (except for the extreme tones), the proposed method achieves around 3,000x shorter runtime and is capable of rendering all tones. These properties further enable the proposed method handling high quality halftones for practical mass printing demands.

The rest of this paper is organized as follows. Section 2 provides an overview of the dot diffusion and its typical feature. Section 3 elaborates the influence of the CT, and Section 4 focuses on the parameter optimization and its influences. Finally, Section 5 presents the simulation results, and Section 6 draws the conclusions.

## 2. DOT DIFFUSION

The concept of the typical dot diffusion as illustrated in Fig. 1 is introduced in this section, where the input grayscale image is of size $P \times Q$. First, the input image is divided into multiple non-overlapped blocks of size $M \times N$ for being processed independently. The processing order ($c[i,j]$, the smaller index value, indicating the earlier processing priority) of each pixel in a block is termed a class matrix (CM). The matrix of a specific size which contains several tiled CMs is termed a grid [11] or a class tiling (CT, $C$) [12]. Normally, the size of a CT can be identical to either the input image [10] or a predefined size, e.g., 256×256 [12]. In the latter case, the CT is periodically tiled to cover the entire image of size $P \times Q$. Notably, the pixels in the image associate to the same $c[i,j] \in C$ can be processed simultaneously to achieve the parallelism property. In the conventional structure [9]-[10], the CMs for all the blocks in an image are identical, and thus induces periodic patterns. This renders an unnatural regularity of the halftone texture. In general, the dot diffusion process of each pixel is formulated as below,
$v[i,j] = x[i,j] + x'[i,j]$, where
$$x'[i,j] = \sum_{m,n} e[i+m, j+m] \times w[m,n]/sum \times H(c[i,j] - c[i+m, j+n]), \quad (1)$$
$e[i,j] = v[i,j] - y[i,j]$, where $y[i,j] = \begin{cases} 255, \text{if } v[i,j] < \gamma \\ 0, \text{if } v[i,j] \geq \gamma \end{cases}$. (2)

In which, $x[i,j] \in [0, L]$ denotes the pixel value of an input image with dynamic range $L$ (=255 for grayscale images); $y[i,j] \in \{0, 255\}$ denotes the binary halftone output; $\gamma = 128$ is suggested in the existing methods [9]-[12]; $w[m,n]$ denotes the coefficient weighting in the diffused matrix (DM) as an example shown in Fig. 2, where in general $\beta \geq \alpha$, and the notation "x" is the central position of the DM with a zero weighting ($w[0,0] = 0$); $H(\cdot)$ denotes the unit step function; term $w[m,n]/sum$ denotes the normalized weighting. Since only the neighboring binarized pixels diffuse $e[i+m, j+m]$ to the current position, the variable $sum$ is the summation of the weightings from those processed pixels as defined below,
$$sum = \sum_{m,n} w[m,n] \times H(c[i,j] - c[i+m, j+n]). \quad (3)$$

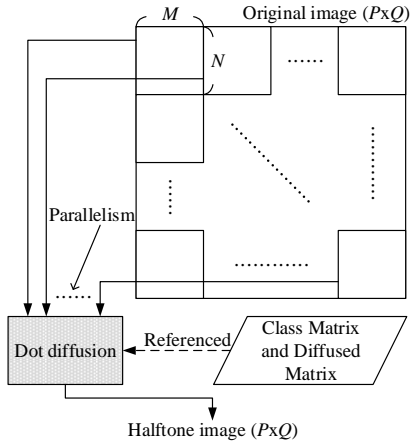

Fig. 1. Traditional dot diffusion flowchart.

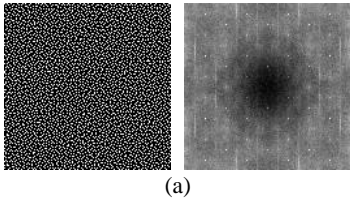

Fig. 2. DM of size 3×3, where identical notation indicates identical weighting.

### 3. CLASS TILING

In this study, we found that the spatial distribution of $c[i,j] \in C$ affects the spectrum property of the generated halftones for dot diffusion methods as shown in Fig. 3. In these two cases, CT is the only difference. In this examination, the averaged power spectrum density (APSD) as that generated with Bartlett's procedure [17] is employed, and it will be detailed in Section 4.2. Figure 3(b) presents a significant improvement in terms of the radial variance, which can be measured by anisotropy. The corresponding two CTs for Figs. 3(a) and 3(b) are shown in Fig. 4. The difference can be fully appreciated via the spatial distribution homogeneity of $c[i,j] \in C$.

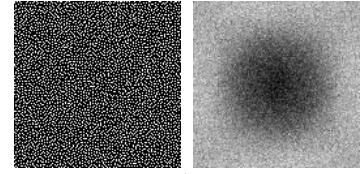

(b)
Fig. 3. Cropped halftones (left) of size 128 × 128 and the corresponding APSDs (right) generated by NADD [12] with (a) their CT and (b) the proposed CT. A constant patch of size 512×512 with grayscale 64 is utilized, and $K = 50$ is applied for the APSD.

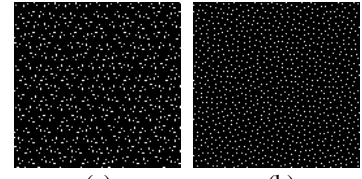

(a) (b)
Fig. 4. Distribution of two different CTs, where $c[i,j] = 0$ presents as white and others are black, and the CM of size 8×8 is supposed. (a) CT in NADD [12]. (b) Proposed CT.

### 3.1. Conventional restriction

Formerly, a CT is constructed by multiple CMs of a fixed size $M \times N$, suggesting that all of the processing orders must appear within each local $M \times N$ spatial region of the CT. Since the same processing pattern is periodically applied to an image, periodic halftone texture is accompanied. This was proved in the analysis of Liu-Guo's work [12] that when a CT containing periodically tiled CMs, a certain periodicity was involved. In addition, the ideal distance among halftone dots [18] with the blue noise property is defined as
$$\lambda_{\bar{g}} = \begin{cases} 1/\sqrt{\bar{g}}, & \text{if } \bar{g} \in [0, 1/4) \\ 2, & \text{if } \bar{g} \in [1/4, 3/4] \\ 1/\sqrt{1-\bar{g}}, & \text{if } \bar{g} \in [3/4, 1] \end{cases}, \quad (4)$$
where $\bar{g} = g/L$, and $g \in [0, L]$ denotes the possible grayscale tone. Thus, to render $g = 1$, the ideal $\lambda_{\bar{g}} \cong 15.97$ in pixels is suggested. When $M < \lambda_{\bar{g}}$, the $g$ cannot be well rendered with a stable distance among dots since the quantization error $e[i,j]$ can only be absorbed by the neighbors with a lower processing priority as defined in Eq. (1). To solve these limitations in the conventional design, each processing order should not be constrained within each $M \times N$ region in a CT. In addition, the positions with the same processing order, i.e., $c[i,j] = 0$ as the case of Fig. 4, are optimized for the preferred spectrum property. In addition, it allows the distances of the positions with the same order $\cong \lambda_{\bar{g}}$ rather than restrained by $M \times N$ as the typical structure.

### 3.2. Distribution control

The iterative halftoning method – DMDBS [13] is employed to render blue noise property, and both homogenous and smooth distribution of the processing orders $c[i,j]$. The corresponding generated result is shown in Fig. 5(a). In their work, the autocorrelation of the point spread function is utilized for simulating the property of Nasanen's HVS model, and it is approximated by a two-component Gaussian kernel as defined below,
$$c_{\tilde{p}\tilde{p}}[m,n] = \frac{180^2}{(\pi D)^2} c_{\tilde{h}\tilde{h}}\left(\frac{180m}{\pi S}, \frac{180n}{\pi S}\right), \text{ where} \quad (5)$$
$$c_{\tilde{h}\tilde{h}}(u,v) = k_1 \exp\left(-\frac{u^2+v^2}{2\sigma_1^2}\right) + k_2 \exp\left(-\frac{u^2+v^2}{2\sigma_2^2}\right). \quad (6)$$
In which, $S = RD$, and $R$ and $D$ denote the resolution in dpi and viewing distance in inch, respectively. In this work, the parameters $(k_1, k_2, \sigma_1, \sigma_2)$ of the two Gaussian models, $\hat{c}_{\tilde{p}_1\tilde{p}_1}[m,n]$ and $\hat{c}_{\tilde{p}_2\tilde{p}_2}[m,n]$, are set at (43.2, 38.7, 0.0219, 0.0598) and (19.1,42.7,0.0330,0.0569), respectively, as determined in Kim-Allebach's work [13] for the best image quality. In addition, the generated dots around that boundary may enlarge the variance of $\lambda_{\bar{g}}$ since the dots are spatially independent during the construction of a CT [4]. Thus, the warp-around property [19], a common trick of building dither array in the field of ordered dithering, is considered to ensure that the dots are spatially dependent for a homogenous texture around CT boundary.

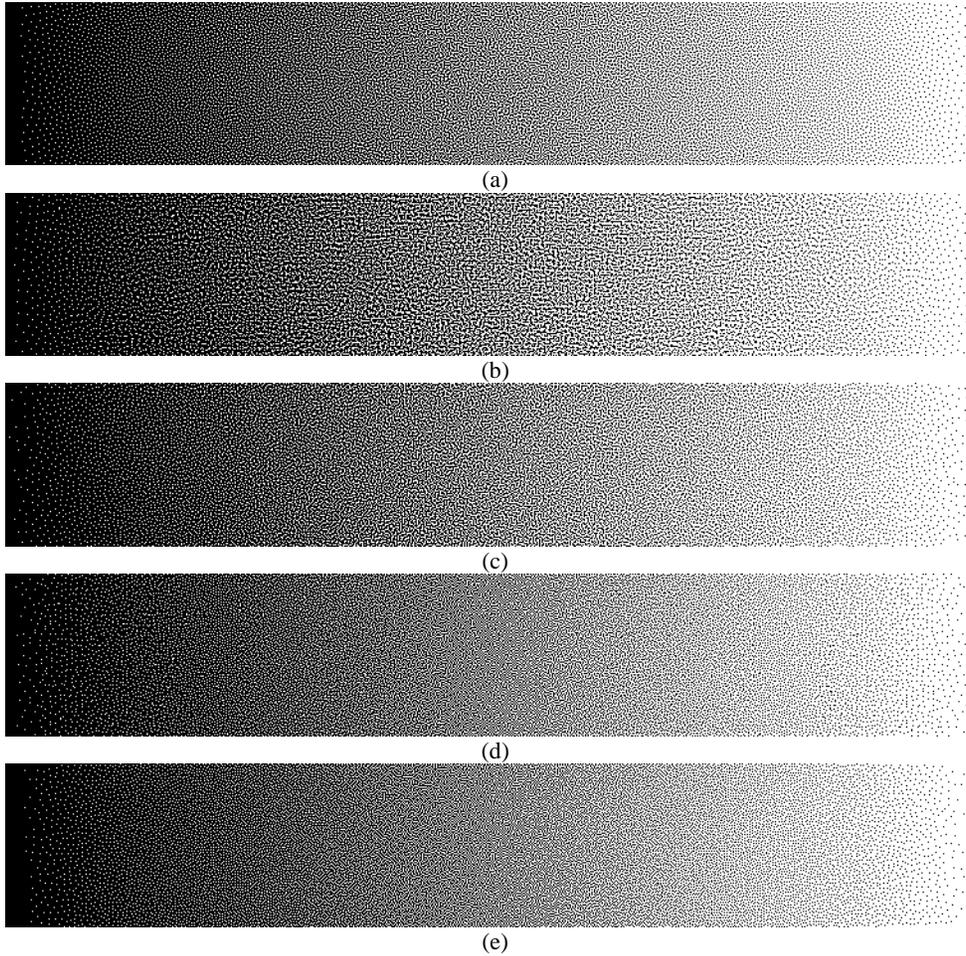

Fig. 5. Ramp halftones of size 768×128. (a) DMDBS [13] with scale parameter $1S$ and (b) $2S$. (c) DD-Pro. (d) DD-NADD [12]. (e) OD-Cha [4].

Although DMDBS generates a great halftone as shown in Fig. 5(a), some extreme tones cannot be rendered since the simulated HVS model is not large enough to capture the sparsity of dots ($\propto \lambda_{\bar{g}}$) which grows rapidly when a tone goes extreme. Specifically, range $0 \leq g \leq 3$ renders no outputs. To control the size of HVS model, the scale parameter ($S$) as defined in Eq. (5) is doubled to enlarge the sampling rate to $\hat{c}_{\tilde{p}_i,\tilde{p}_i}(x,y)$. The corresponding ramp result is shown in Fig. 5(b). Although it renders the extreme tones, granules appear at midtone areas. To have an in-depth exploration, the cases of extreme tones are exhibited in Table I. It shows that even though randomized textures appeared at midtone area with $2S$, performance at extreme area is still quite stable as that with unadjusted scale parameter ($1S$). In our case, models $\hat{c}_{\tilde{p}_i,\tilde{p}_i}[m,n]$ with $1S$ and $2S$ are used for tones $4 \leq g \leq 251$ and the rest tones, respectively.

### 3.3. CT construction

To obtain a CT, masks $\{I_g\}_{g=0}^{L}$ are successively designed from 0 to $L$ by the DMDBS, where each mask $I_g[m,n] \in \{0,1\}$ equals to $y[m,n]$ with the input $x[m,n] = (L-g)/L$ as defined in Section 3.2. During the process, the stacking constraint, $I_g[m,n] = 0$ if $I_{g-1}[m,n] = 0$, is applied. Subsequently, the prototype of CT ($F$) is constructed as

$$f[m,n] = \begin{cases} g, \text{ if } I_g[m,n] = 0 \land I_{g-1}[m,n] = 1 \\ 0, \text{ O.W.} \end{cases} \quad (7)$$

To maintain the parallelism of the typical dot diffusion, the CT can be formed with the given CM size from quantizing $F$ as

$$c[m,n] = \lfloor f[m,n] \times (M \times N)/(L+1) \rfloor, \quad (8)$$

where $L$ denotes the maximum tone value; $M \times N$ denotes the CM size, and $\lfloor \cdot \rfloor$ denotes the floor operation. Figure 4(b) shows an example of the constructed CT. Thus, only $M \times N$ runtime units are needed for the entire image halftoning process when required number of threads are deployed.

## 4. OPTIMIZATION

All the remaining parameters of the proposed dot diffusion are optimized to substantially improve halftone quality. However, some potential issues are involved with the use of cost functions and the well-known Bartlett's procedure [17] during optimization. These issues are discussed in this section.

### 4.1. Cost functions

In general, a cost function is defined to evaluate the difference between the generated halftone pattern and an expected output. To this end, the perceived error was individually utilized with a HVS-like model for a homogenous halftone texture and better similarity to the tones of interest [12], [20]-[21]. In addition, the power spectrum density (PSD) is employed to measure whether the blue noise property is met [6]-[7], [21]-[23]. For this, Zhou and Fang [6] evaluated the correlations of the three various directions on PSD for a circle-shape spectrum, and the power concentricity was estimated through the power ratio between the entire spectrum and those under the cutoff frequency. In addition, a more intuitive way is to calculate the PSD difference between the generated halftones and the ground truth. In Li and Allebach's work [21], the visually weighted root-mean-squared error was minimized for both highlight and shadow regions. The mean-square error in the midtone area between the magnitudes of the direct binary search (DBS) and a halftone output was defined as

$$\varepsilon = \sum_{k,l}\left(\hat{G}'[k,l] - \hat{G}^I[k,l]\right)^2, \quad (9)$$

where $\hat{G}'[k,l]$ and $\hat{G}^I[k,l]$ denoted the estimated magnitudes of the halftone output and the ideal DBS, respectively. In Chang and Allebach's work [22], a single cost function was utilized for all grayscales with the averaged PSDs (APSDs) rather than the above magnitudes. In addition, the cost function was normalized with the spectrum of the DBS for handling their variances as formulated below,

$$\varepsilon = \sum_{k,l} \frac{\left(\hat{g}'[k,l] - \hat{g}^I[k,l]\right)^2}{\hat{g}^I[k,l]^2}, \quad (10)$$

where $\hat{g}'[k,l]$ and $\hat{g}^I[k,l]$ were the estimated APSDs obtained from an evaluated halftone image and the one generated by the DBS, respectively. However, this normalization term endows the cost at a lower frequency with a higher weighting to dominate the entire estimated cost. In Han et al.'s work [23], the normalization term was modified as

$$\varepsilon = \sum_{k,l} \frac{\left(\hat{g}'[k,l] - \hat{g}^I[k,l]\right)^2}{\hat{g}'[k,l]^2 + \hat{g}^I[k,l]^2}. \quad (11)$$

TABLE I. DMDBS Results of Size 128×128 with Different Scale Parameters ($S$) and Corresponding PSDs [17]. Results of $1S$ at $g = [1,3]$ Are Not Shown since They Render No Dots.

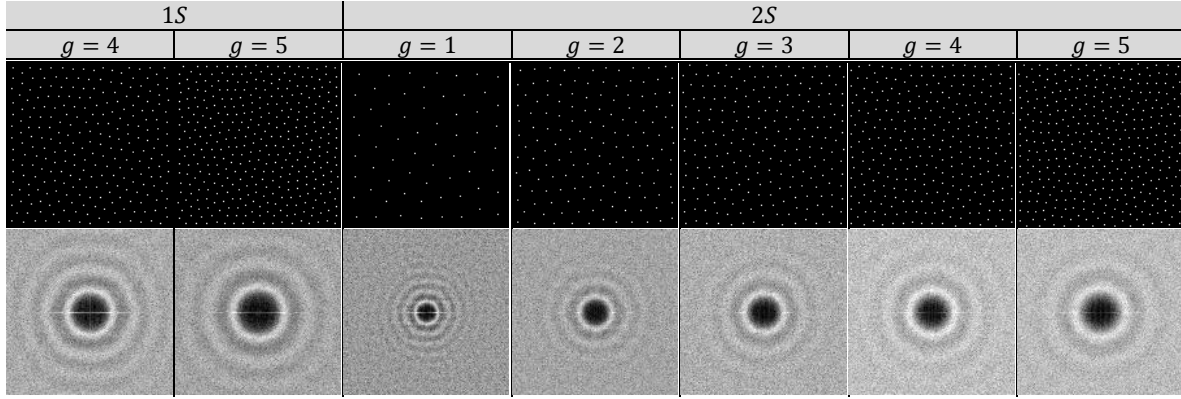

| $1S$ | | $2S$ | | | | |
|---|---|---|---|---|---|---|
| $g = 4$ | $g = 5$ | $g = 1$ | $g = 2$ | $g = 3$ | $g = 4$ | $g = 5$ |

TABLE II. Averaged Power Spectrum Densities at Grayscale 16. Window Size is Set at 128×128.

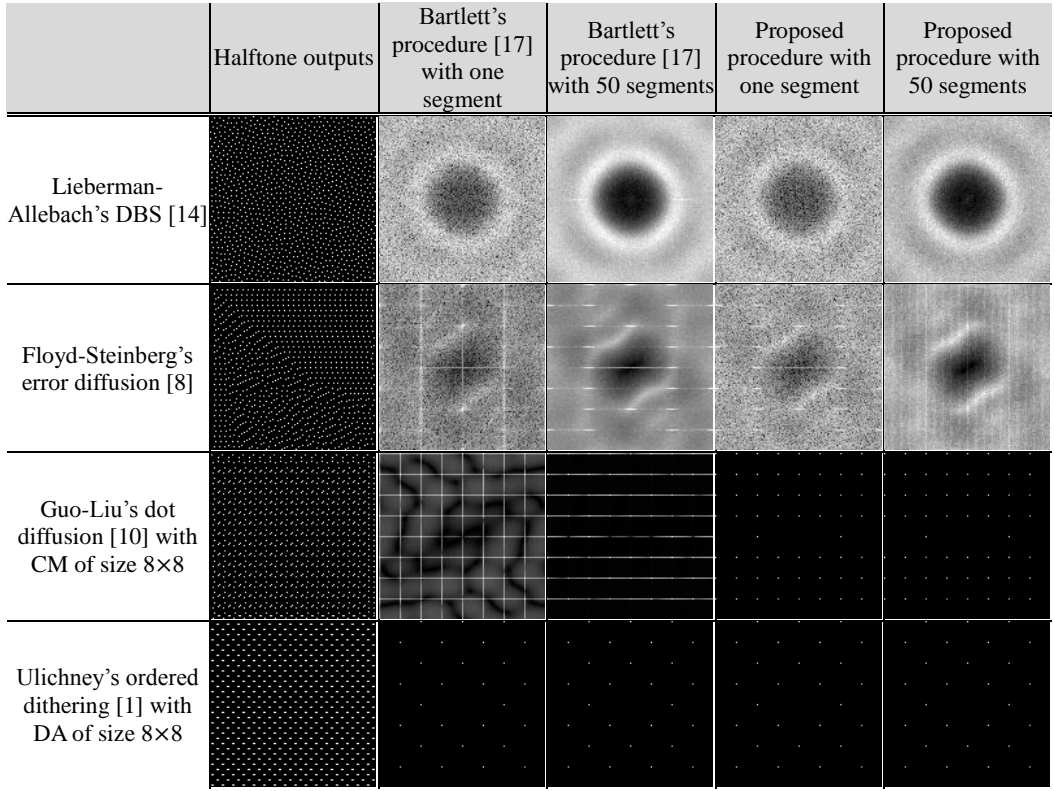

| | Halftone outputs | Bartlett's procedure [17] with one segment | Bartlett's procedure [17] with 50 segments | Proposed procedure with one segment | Proposed procedure with 50 segments |
|---|---|---|---|---|---|
| Lieberman-Allebach's DBS [14] | | | | | |
| Floyd-Steinberg's error diffusion [8] | | | | | |
| Guo-Liu's dot diffusion [10] with CM of size 8×8 | | | | | |
| Ulichney's ordered dithering [1] with DA of size 8×8 | | | | | |

This cost function evaluates the weighted cost evenly over all frequencies. As introduced above, currently two types of cost functions are presented for different purposes: 1) Perceived error: it evaluates the visual signal similarity, and it cannot reflect the property of dot distribution or even the similarity to the blue noise spectrum;

2) spectrum error: it has a complement property to the perceived error. Former methods considered one type of cost function for each tone for optimization. However, the independently used spectrum error may encounter an identical ground truth as that defined in Eq. (4): $\lambda_{\bar{g}} = 2$ even though they are rendering different tones. This issue raises when an optimization involves a factor which affects the dot density, and it ends up with an identical density halftone for different grayscales. In this work, the perceived error is also considered to maintain the correct proportion of dot density on different tones.

**4.2. Averaged power spectrum density (APSD)**
Bartlett's procedure [17] is a well-known spectral analysis for halftoning techniques and it is first used in Ulichney's work [1] for halftone analysis. It averages periodograms of many short divided segments from an available signal to yield a zero variance result. A one dimensional example can be formulated as
$$q_r[n] = q[rR + n]w[n], \text{ where } 0 \leq n \leq M - 1, \quad (12)$$
where $q_r[n]$ denotes the $r$-th segment of the signal $q[n]$; $w[n]$ denotes a window of size $M$ (in two dimensional case, a rectangular window of size $M \times N$ is utilized); $R$ denotes the step size of each segment. Supposing that $I_r(\omega)$ is the periodogram of $q_r[n]$, the averaged periodogram is defined as
$$\bar{I}(\omega) = \frac{1}{K}\sum_{r=0}^{K-1} I_r(\omega). \quad (13)$$
In general, $R = M$ can be reasonably assumed for a continuous and non-overlapped sampling since the segments are considered as independent and identically distributed (i.i.d.) random variables [17]. This assumption holds true when the positions of halftone dots lean to zero cross-correlation, e.g., the outputs generated by iterative halftoning methods and error diffusion methods. However, it cannot be endorsed when a halftone pattern is suffered from the periodic artifact, in particular when $R$ is fully divided by its periodicity, and ends up with a biased property. A concrete case is shown in the results of Guo-Liu's dot diffusion [10] estimated with Bartlett's procedure [17] in Table II, where the window of size 128×128 is fully divided by the periodicity of 8×8, and a vertical and continuous ($R = M$) sampling is used. Herein, all of the APSDs are averaged with $K$ independent segments from a halftone pattern of size 128×(128× $K$). In this experiment, $K = 1$ and 50 are supposed for the unstable and stable results, respectively. In which, $K > 50$ will have a saturated output as that of $K = 50$. It is clear that when $K = 50$, Bartlett's procedure only shows the horizontal periodicity (vertically spaced lines).

In addition, Ulichney [24] suggested that the windows should be located far from the boundary or the edge of an available signal to capture the "steady-state" segments to avoid the transient effect as represented as the horizontal line appeared in the DBS's averaged estimation shown Table II. The transient effect usually shows up around the edge of a halftone pattern. However, the suggested locations far from the boundary may occasionally meet the periodicity of a certain halftone patterns, and thus also end up with a biased property.

To avoid the potential biased property and transient effect as indicated above, an alternative randomly overlapped sampling method is proposed. The overlapping strategy was adopted in Welch's work [25] with $R = M/2$, and it further reduced the variance of the averaged periodogram by almost a factor of two for a fixed amount of signal because this doubles the number of segments. Notably, the increase of the segment number does not continue to reduce the variance since the segments become more dependent along with the increase of overlapped area [17]. To avoid the cases that the sampled segment equals to the periodicity of the halftone pattern, a random sampling within a halftone pattern of a given size is utilized. In our case, $K = 50$ segments $q_r[n]$ are randomly captured by a window of size 128×128 within a halftone image of size 512×512 with a constant tone, and this image size is greater than the periodicity of the evaluated halftone patterns. In addition, the $I_r(\omega)$ defined in Eq. (13) is generated by the discrete Fourier transform (DFT) from $q_r[n]$. Table II shows the corresponding $\bar{I}(\omega)$ as defined in Eq. (13). Notably, only one segment of $q_r[n]$ is shown because of the limited pages. Comparing with Bartlett's procedure, the estimates of both DBS and ordered dithering (OD) which barely have the transient effect show a similar property as that of the proposed procedure. A slight difference can be found by comparing with both of the 50 averaged periodograms of the DBS. The proposed procedure further eliminates the slight transient effect as represented as a horizontal line shown in Bartlett's result. Moreover, the proposed procedure offers a more unbiased property to the ones which have either transient effect (error diffusion) or periodicity artifact (dot diffusion), in particular the periodicity of the dot diffusion pattern in terms of both horizontal and vertical directions are both presented in the estimate with the proposed procedure. For the case of the error diffusion, the proposed procedure fairly and proportionally reflects the property. The error diffusion pattern is over-enhanced as the left hand side regular dot distributions. The proposed procedure is utilized in our optimization procedure.

**4.3. Algorithm**
Formerly, the error diffusion weighting ($w[m,n]$) and the threshold ($\gamma[i,j]$) as defined in Eqs. (1)-(2) were both demonstrated with high dependency to the input tones [21]. Since the CT has been proved of significant effect on the spectrum property as discussed in Section 3, $w[m,n]$ and $\gamma$ in Eqs. (1)-(3) are replaced with $w[m,n;x[i,j],f[i,j]]$ and $\gamma(x[i,j],f[i,j])$, respectively, in the proposed dot diffusion. In which, the weighting is further subject to $\alpha(x[i,j],f[i,j]) \geq 0$ and $\beta(x[i,j],f[i,j]) \geq 0$ according to Fig. 2; $f[i,j] \in F$ is defined in Eq. (7). In addition, in contrast to the former tone-dependent works [7], [21], the additional order-dependent design exploits the expected spectrum property of CT distribution as introduced in Section 3. Consequently, for each grayscale $g$ and unquantized processing order $f$, a three-dimension vector $\{\alpha(g,f), \beta(g,f), \gamma(g,f)\}$ is needed to be optimized with the following algorithm.

**Parameter Optimization Algorithm**
**Variable.**
$g \leq L$: Grayscales.
$f \leq L$: Unquantized processing order.
$\mathcal{H}_{g,f} = \{\alpha(g,f), \beta(g,f), \gamma(g,f)\}$
$$\mathcal{P} = \begin{bmatrix} \mathcal{H}_{0,0} & \cdots & \mathcal{H}_{0,L} \\ \vdots & \ddots & \vdots \\ \mathcal{H}_{127,0} & \cdots & \mathcal{H}_{127,L} \end{bmatrix}$$
**Begin stage 1.**
1. Initialize $\forall \mathcal{H}_{g,f} \in \mathcal{P}$ with $\{0,0,f\}$.
For $g = 0$ to 127
   2. Initialize $k \leftarrow 0$ and $e_{opt} \leftarrow \infty$, where $k$ denotes $k$-th iteration and $e_{opt}$ denotes the optimum error.
   3. For each $f$, obtain $\mathcal{H}'_{g,f}$ to yield the minimum spectrum error $e'_{f_k}$ with Eq. (11) through the downhill search algorithm. Notably, only $\mathcal{H}'_{g,f}$ at the evaluating $f$ is modified for each $e'_{f_k}$, and other $\mathcal{H}_{g,f}$ in $\mathcal{P}$ remain the same. For instance, $e'_{0_k}$ is derived with $[\mathcal{H}'_{g,0}, \mathcal{H}_{g,1}, \ldots, \mathcal{H}_{g,L}]$.
   4. Replace $\mathcal{H}_{g,f} \in \mathcal{P}$ with the $\mathcal{H}'_{g,f_k^*}$ and go back to Step 3 with $k \leftarrow k+1$ and $e_{opt} \leftarrow e'_{f_k^*}$ if $e'_{f_k^*} < e_{opt}$, where
   $$f_k^* = \underset{f_k}{\arg\min}(e'_{f_k}). \quad (14)$$
   Otherwise, go back to Step 2 with $g \leftarrow g+1$, and a new $\mathcal{P}'$ is obtained.
**Begin stage 2.**

For $g = 0$ to 127
  5. Perform the same initialization as Step 2.
  6. For each $f$, identical process as Step 3 is performed based upon the $\mathcal{P}'$ with the perceived error, which is measured by the human-visual mean-square-error (HMSE) defined in Eq. (16) in the next section.
  7. Perform the same process as Step 4, and the new $\mathcal{P}_{opt}$ is applied to the proposed dot diffusion after all the iterations.

**End.**

The above algorithm obtains the parameters for $g < 128$, and the optimized $\{\alpha(L-g,f), \beta(L-g,f), \gamma(L-g,f)\}$ is also applied for $g \geq 128$. In this algorithm, Eq. (11) is first applied in stage 1 for each tone for a high similarity to the expected spectrum property. In addition, the PSD generated from the DMDBS [13] is adopted as the ground truth as opposed to the methods [23] which previously utilized DBS. The downhill search algorithm is employed with the optimized $\mathcal{P}'$ for a lower perceived error for each tone to avoid the issue raised at the end of Section 4.1. Since the downhill search pursuits the local optimum, the $\mathcal{P}_{opt}$ can yield a great balance between the spectrum error and perceived error. Moreover, initializing $\gamma(g,f) = f$ in Step 1 treats the threshold distribution as the cutting-edge dither array design [4] for a good spectrum property at the early stage. In the same step, both $\alpha(g,f)$ and $\beta(g,f)$ are initially set to zero for simulating the results of ordered dithering (simply thresholding operation is used). This setting additionally benefits the processing efficiency since there is a chance of no further reduction on the cost based upon the initial values. This suggests that if these two parameters equal to zero in $\mathcal{P}_{opt}$, there is no need to diffuse the quantization error, and thus saves the computation time.

## 5. SIMULATION RESULTS
### 5.1. Comparison across various categories
Eight halftoning methods of various types are adopted for comparison. Herein, the iterative-based methods, error diffusion, dot diffusion, and ordered dithering are abbreviated as IT, ED, DD, and OD, respectively. These methods and the related settings are defined as follows: 1) DMDBS [13] (abbr.: IT-DMDBS), 2) Ostromoukhov's ED [7] (abbr.: ED-Ost), and 3) Zhou-Fang's ED [6] (abbr.: ED-Zho). For dot diffusion, the CM of size 8×8 is considered for comparison, including: 4) NADD [12] (abbr.: DD-NADD), 5) Guo-Liu's DD [10] (abbr.: DD-Guo), and 6) the proposed dot diffusion (abbr.: DD-Pro). In addition, two ordered dithering methods are compared as well: 7) Chandu et al.'s method [4] (abbr.: OD-Cha): the binary version is used, and 8) Kacker-Allebach's OD [5] (abbr.: OD-Kac): four screens of size 32×32 are used. Notably, the CT size of both DD-NADD and DD-Pro and the screen size of OD-Cha are set at 256×256 for a fair comparison.

In terms of the image quality, the human-visual peak signal-to-noise ratio (HPSNR) [26] is utilized for evaluation as formulated below,

$$\text{HPSNR} = 10\log_{10}\left(\frac{255^2}{\text{HMSE}}\right), \text{ where} \quad (15)$$

$$\text{HMSE} = \frac{1}{P \times Q}\sum_{i=1}^{P}\sum_{j=1}^{Q}\left[\sum_{m,n} w[m,n](x[i+m,j+n] - y[i+m,j+n])\right]^2. \quad (16)$$

The variables $x[i,j]$ and $y[i,j]$ follow the definitions of Eqs. (1)-(2); $P \times Q$ denotes the image size; $w[m,n]$ denotes the weighting to simulate the lowpass characteristic property of the human visual system. Normally, the kernel size is determined by the viewing distance and resolution (dpi) [27], and the number of pixels in one visual degree can be modeled with the following formula,

$$N_v = r \times R \times \frac{cm}{inch}, \text{ where } r = 2 \times D \times \tan\left(\frac{\theta}{2}\right). \quad (17)$$

Herein, $\theta = 1°$ denotes the viewing degree; $D$ denotes the viewing distance in centimeters (cm); $R$ denotes the image resolution in dpi; $r$ denotes the viewed width, and $cm/inch = 0.393700787$. To cover the most of the configurations in viewing halftone images, two frequently used viewing distances, 15 cm and 30 cm, and resolutions, 75 dpi and 150 dpi, are involved for a complete comparison. Thus, totally three Gaussian kernels of sizes 7×7, 15×15, and 31×31 are adopted for a fair evaluation.

Figure 6 shows the corresponding performances, in which each method has three points for their HPSNR with different Gaussian kernel sizes, and the greater size obtains a higher HPSNR. Each HPSNR is averaged from the results of 254 single-tone images of size 512×512 within grayscale range $g \in [1,254]$. For the runtime, the simulation platform is with a 32GB RAM and a 3.4GHz CPU which is equipped with eight threads. Notably, although multiple threads are supported, only both DD and OD can be further speeded up by considering their parallel algorithms. The shown runtimes reflect the properties of the halftoning methods in terms of their processing complexity: IT requires the longest runtime for its inherent iteration approach, and the OD can obtain the fastest speed by their simple thresholding process. Although both ED-Ost and ED-Zho have fewer numbers of diffused neighbors than that of the DD methods, more runtime is required on the two ED methods since the parallelism is not available. This figure also shows that the proposed DD-Pro is faster than other DD methods, since there is no need to diffuse error when both $\alpha$ and $\beta$ are equal to zero as discussed in Section 4.3. On the other hand, ED methods have the best image quality in terms of the HPSNR, and DD sacrifices a bit on image quality with the trade-off on its parallelism advantage. Normally, OD has a relatively low image quality because it cannot compensate the quantization error from the neighboring pixels. Yet, the OD-Cha has a good performance by enjoying its stochastic dispersed halftone texture. In accordance with the above analysis, the DD-NADD, OD-Cha, and the proposed DD-Pro have a great superiority in terms of the HPSNR compared to other methods, and these methods also have the additional parallelism feature. These methods are further compared in detail in the following subsection.

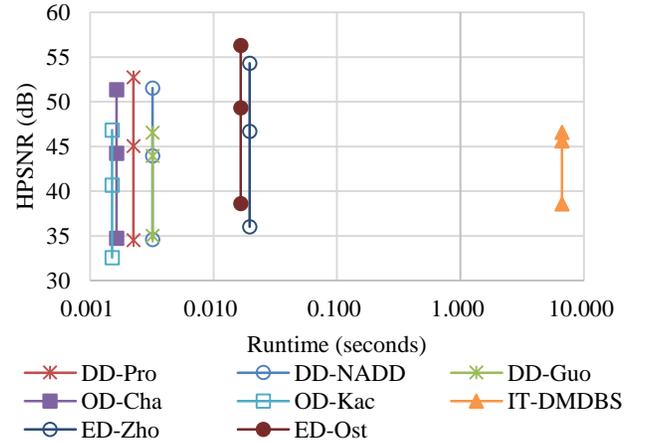

Fig. 6. Image quality and runtime of various methods, where the nodes of each method from bottom to top indicate the average HPSNRs with kernels of sizes 7×7, 15×15, and 31×31, respectively.

### 5.2. Halftone textures
This section further explores the visual quality of the halftone results. Figure 5(c)-(e) shows the ramp halftones of the DD-Pro, DD-NADD, and OD-Cha, where Fig. 5(a) can be regarded as the result with an ideal blue noise distribution for comparison. As is can be seen, both DD-NADD and OD-Cha have an obvious transient effect [6] around $g = 128$. It appears around the dramatic changes on the density of rendered dots, and thus introduces the density discontinuity. Looking at $g = 64$ and 192, both of the DD-NADD and OD-Cha render a weak homogeneity. In addition, DD-NADD presents a noisy texture, and OD-Cha has plenty of horizontal and vertical artifacts. Thus, both of them show a weak smoothness at the same locations. At the areas of around $g = 0$ and 255, DD-NADD is the only method introduces the worm artifact which reduces its visual quality. In contrast to the other two methods, the proposed DD-Pro has a prominent superiority in terms of both the homogeneity and smoothness at each grayscale.

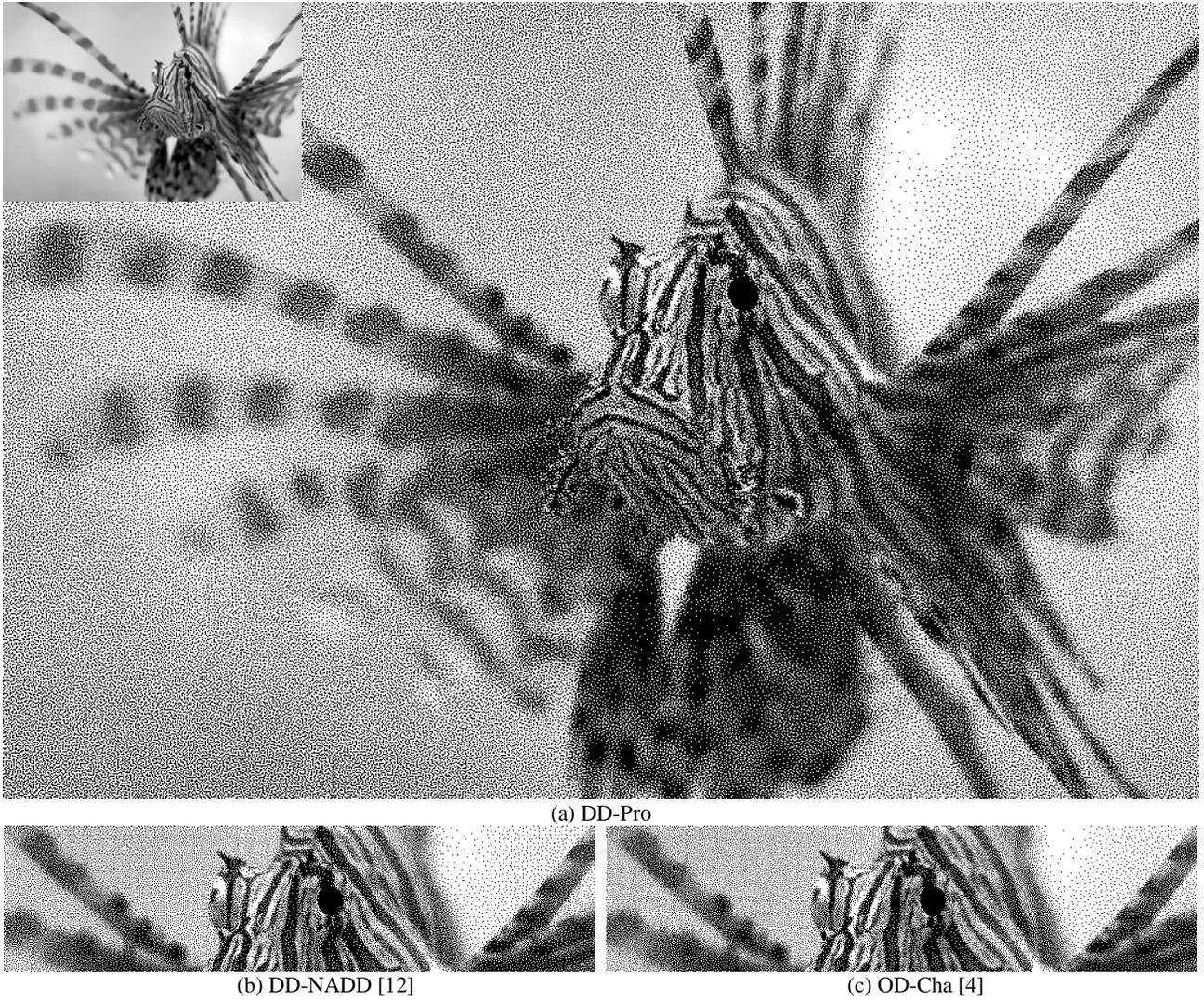

Fig. 7. Result of the test image, *Lion Fish* of size 1024×683 (License: Celeste RC, flickr.com, CC BY-NC), where the top-left corner of subfigure (a) shows the original image.

Figure 7 shows the rendered outputs with a natural image. This image has a high contrast, fine structure, and various spatial frequencies and flat regions of dark and bright colors, thus it is a good benchmark to demonstrate the halftoning performance in terms of natural image rendering. Figure 7(a) shows the output of the proposed method. Apparently, no blocking effect is involved even it is processed by multiple periodically tiled CTs. Figure 7(b)-(c) shows the comparison among various methods with specific cropped parts. As it can be seen, the inhomogeneous backgrounds seriously lead to noisy perception on DD-NADD and OD-Cha, and thus they are of lower visual quality in contrast to that of the proposed method.

**5.3. Power spectrum density**

Another point of view is examined to have an in-depth and concrete exploration on the homogenous property as discussed in the previous section. To that end, the modified APSD as presented in Section 4.2 is considered as the metric. Table III shows the corresponding results and halftone patches at various constant grayscales. It is noteworthy that both of the DD-NADD and OD-Cha cannot render dots in extreme grayscale areas (labeled "n/a"). Comparing with the DD-NADD, it is obviously that the proposed method shows no periodical artifact which is normally represented as certain impulse power dots on APSD as in the case of DD-NADD. These periodical impulse dots in DD-NADD's results reveal the periodicity of its halftone patterns. Yet, the power spectra of the DD-Pro do not have this artifact since it avoids the limitation of the conventional CM as discussed above. Although the memory requirement is increased for the entire CT of the proposed method, only 67KB (=256×256×1 bytes for CT + 256×3×4 bytes for $\mathcal{P}_{opt}$) is required. It cannot be a big issue for the currently modern devices. In addition, the noisy power spectrum of the DD-NADD indicates the unstable grayscale rendering capability, which concretely embodies with the various densities of halftone dots as the case at $g = 8$.

For the comparison with the OD-Cha, no major difference is shown when $g \leq 16$ according to the results of Table III. However, the power spectrum of OD-Cha can be further classified into three groups when $g \geq 32$ Herein, each group is separately by the two circles with two different colors. This phenomenon is caused by the two types of halftones with different frequencies are used to construct the halftone patterns of the OD-Cha. An extreme case is shown at $g = 128$, which contains stochastic dispersed texture as that of the DD-Pro and the chessboard structure shown in DD-NADD's result, simultaneously. The introduction of the difference between the two types of different textures also induces a discontinuous dot density, termed transient effect, as shown in the ramp image result and natural image output of Figs. 5(e) and 7(c), respectively. Conversely, the stability of the proposed method totally avoids the discontinuous textures, and thus generates a homogenous texture over all grayscales.

TABLE III. HALFTONE RESULTS OF SIZE 128×128 AND CORRESPONDING APSD, WHERE N/A DENOTES RENDERING NO DOTS.

| | $g = 1$ | $g = 2$ | $g = 4$ | $g = 8$ | $g = 16$ | $g = 32$ | $g = 64$ | $g = 128$ |
|---|---|---|---|---|---|---|---|---|
| DD-Pro | | | | | | | | |
| DD-NADD [12] | n/a | | | | | | | |
| OD-Cha [4] | n/a | n/a | n/a | | | | | |

### 5.4. Discussions

Although the proposed method obtains a bit lower image quality comparing to the former error diffusion methods as shown in Fig. 6, around 8x faster speed can be provided when eight threads are available. Although the IT-DMDBS can yield the highest image quality as it can be seen in Fig. 6, it cannot well render the extreme tones as introduced in Table I. In addition, the runtime is much longer than that of the proposed method by a factor of around 3,000 (=6.65368/0.00224). In addition, according to the experimental results, the proposed method achieves the best visual image quality among the scope of all the state-of-the-art halftoning methods with parallelism as shown in Fig. 6. Other evidences can be seen from the homogenous halftone texture as shown in Figs. 5 and 7, and the artifact-free property demonstrated in Fig. 5 and Table III.

In terms of the processing structure, the ordered dithering simply applies thresholding for halftoning, while the proposed method additionally accompanies the advantage of error diffusion to compensate regional tone. Although the proposed method presents a bit slower speed by about 1.37x (=0.00224/0.00164 as shown in Fig. 6) to the OD-Cha, a more stable and accurate tone presentation capability, and artifact-free property are both endorsed. Figure 5 and Table III demonstrate the identical observation. In addition, the dots generated by the former dot diffusion cannot accurately present each tone since the same processing order $c[i,j]$ and threshold $\gamma$ in CT have no spatial relation, and thus it is difficult to render a stable dot density as shown in Table III. A summary of performance is organized in Table IV, where the "image quality" is the average of HPSNR with three different kernel sizes as shown in Fig. 6; "speed" is identical to Fig. 6; "periodicity" is determined by the utilized CT for both DD methods or the dither array for the OD method; the artifacts as listed in the last two columns are quantized for comparison. Notably, the IT-DMDBS is involved for comparison as an iterative halftoning method.

### 6. CONCLUSIONS

Formerly, ordered dithering mainly focuses on the threshold arrangements, and dot diffusion is implemented with the omnidirectional error diffusion as oppose to the typical error diffusion methods which diffuse the errors to specific orientations. In this study, the proposed dot diffusion utilizes the advantages from both ordered dithering and dot diffusion for a great visual quality and high processing efficiency. In addition, the proposed method enhances the spatial relationship among the processing orders in CT to significantly improve the homogeneity and smoothness of halftones. Specifically, an alternative approach on APSD calculation as opposite to the typical

TABLE IV. SUMMARY OF COMPARISON WITH STATE-OF-THE-ARTS (THE BEST VALUE IN EACH CATEGORIES IS CIRCLED).

| Methods | Image similarity (dB) | Speed (seconds) | Periodicity (pixels) | Extreme value rendering | Transient effect | Chessboard texture |
|---|---|---|---|---|---|---|
| DD-Pro | 44.1 | 0.00224 | 256 | $g = \{\cdot\}$ | No | No |
| IT-DMDBS [13] | 43.6 | 6.65368 | ∞ | $g = \{1, ..., 3, 252, ..., 254\}$ | No | No |
| DD-NADD [12] | 43.4 | 0.0032 | 256 | $g = \{1, 254\}$ | Yes | Yes |
| OD-Cha [4] | 43.4 | 0.00164 | 256 | $g = \{1, ..., 4, 251, ..., 254\}$ | Fair | Fair |

Bartlett's procedure is proposed to correctly reflect the property of halftone patterns. This approach is a good tool to highlight the periodic artifact of the halftone patterns. As documented in the experimental results, the proposed dot diffusion is substantially superior to the former dot diffusion and ordered dithering in terms of visual quality. Although the runtime of the proposed method is slightly slower than that of the cutting-edge OD, the proposed method with artifact-free property offers a great market potential. In contrast to those methods which do not offer parallelism property, the proposed method meets the demand of the practical industries. Particularly, the increasing on image resolution requires highly efficient processing and mass productivity. The proposed scheme can be a very good candidate to address these issues.

## REFERENCES

[1] R. Ulichney, *Digital halftoning*, Cambridge, MA: MIT Press, 1987.
[2] J. B. Rodriguez, G. R. Arce, and D. L. Lau, "Blue-noise multitone dithering," *IEEE Trans. Image Processing*, vol. 17, no. 8, pp. 1368-1382, Aug. 2008.
[3] D. L. Lau and R. Ulichney, "Blue-noise halftoning for hexagonal grids," *IEEE Trans. Image Processing*, vol. 15, no. 5, pp. 1270-1284, May 2006.
[4] K. Chandu, M. Stanich, C. W. Wu, and B. Trager, "Direct multi-bit search (DMS) screen algorithm," in *Proc. IEEE ICIP*, pp. 817-820, 2012.
[5] D. Kacker and J. P. Allebach, "Aperiodic micro screen design using DBS and training," in *Proc. SPIE - The International Society for Optical Engineering*, vol. 3300, pp. 386-397, 1998.
[6] B. Zhou and X. Fang, "Improving mid-tone quality of variable coefficient error diffusion using threshold modulation," *ACM Trans. on Graphics*, vol. 22, no. 3, pp. 437-444, July 2003.
[7] V. Ostromoukhov, "A simple and efficient error-diffusion algorithm," in *Proc. SIGGRAPH*, pp. 567-572, 2001.
[8] R. W. Floyd and L. Steinberg, "An adaptive algorithm for spatial gray scale," in *Proc. SID 75 Digest. Society for information Display*, pp. 36-37, 1975.
[9] M. Mese and P. P. Vaidyanathan, "Optimized halftoning using dot diffusion and methods for inverse halftoning," *IEEE Trans. Image Processing*, vol. 9, no. 4, pp. 691-709, April 2000.
[10] J. M. Guo and Y. F. Liu, "Improved dot diffusion by diffused matrix and class matrix co-optimization," *IEEE Trans. Image Processing*, vol. 18, no. 8, pp. 1804-1816, Aug. 2009.
[11] S. Lippens and W. Philips, "Green-noise halftoning with dot diffusion," in *Proc. SPIE - The International Society for Optical Engineering*, 2007.
[12] Y. F. Liu and J. M. Guo, "New class tiling design for dot-diffused halftoning," *IEEE Trans. Image Processing*, vol. 22, no. 3, pp. 1199-1208, March 2013.
[13] S. H. Kim and J. P. Allebach, "Impact of HVS models on model-based halftoning," *IEEE Trans. Image Processing*, vol. 11, no. 3, pp. 258-269, March 2002.
[14] D. J. Lieberman and J. P. Allebach, "Efficient model based halftoning using direct binary search," in *Proc. IEEE International Conference on Image Processing*, vol. 1, pp. 775-778, Oct. 1997.
[15] C. Schmaltz, P. Gwosdek, A. Bruhn, and J. Weickert, "Electrostatic halftoning," *Computer Graphics*, vol. 29, pp. 2313-2327, 2010.
[16] D. E. Knuth, "Digital halftones by dot diffusion," *ACM Trans. Graph.*, vol. 6, no. 4, pp. 245-273, Oct. 1987.
[17] A. V. Oppenheim and R. W. Schafer, *Discrete-Time Signal Processing*, 2$^{nd}$, Englewood Cliffs, NJ: Prentice-Hall, 1999.
[18] G. J. Garateguy and G. R. Arce, "Voronoi tessellated halftone masks," in *Proc. IEEE ICIP*, pp. 529-532, Sept. 2010.
[19] B. W. Kolpatzik and J. E. Thornton, "Image rendering system and method for generating stochastic threshold awways for use therewith," U.S. Patent 5 745 660, Apr. 28, 1998.
[20] P. W. Wong, "Adaptive error diffusion and its application in multiresolution rendering," *IEEE Trans. Image Processing*, vol. 5, no. 7, pp. 1184-1196, July 1996.
[21] P. Li and J. P. Allebach, "Tone-dependent error diffusion," *IEEE Trans. Image Process.*, vol. 13, no. 2, pp. 201–215, Feb. 2004.
[22] T. Chang and J. P. Allebach, "Memory efficient error diffusion," *IEEE Trans. Image Processing*, vol. 12, pp. 1352–1366, Nov. 2003.
[23] S. W. Han, M. Jain, R. Kumontoy, C. Bouman, P. Majewicz, and J. P. Allebach, "AM/FM halftoning: improved cost function and training framework," in Proc. SPIE, Color Imaging XII: Processing, Hardcopy, and Applications, vol. 6493, Jan 2007.
[24] R. A. Ulichney, "Dithering with blue noise," in *Proc. IEEE*, vol. 76, no. 1, pp. 56-79, 1988.
[25] P. D. Welch, "The use of the fast Fourier transform for the estimation of power spectra," *IEEE Trans. Audio Electroacoustics*, vol. AU-15, pp. 70-73, June 1970.
[26] J.-M. Guo and Y.-F. Liu, "Improved block truncation coding using optimized dot diffusion," *IEEE Trans. Image Processing*, vol. 23, no. 3, pp. 1269-1275, March 2014.
[27] J.-M. Guo and Y.-F. Liu, "Joint compression/watermarking scheme using majority-parity guidance and halftoning-based block truncation coding," *IEEE Trans. Image Processing*, vol. 19, no. 8, pp. 2056-2069, Aug. 2010.

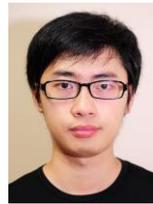

**Yun-Fu Liu** (S'09-M'13) received the master's degree in electrical engineering from Chang Gung University, Taoyuan, Taiwan, in 2009, and the Ph.D. degree in electrical engineering from the National Taiwan University of Science and Technology, Taipei, Taiwan, in 2013.

He was involved in research with the Department of Electrical and Computer Engineering, University of California at Santa Barbara, Santa Barbara, in 2012. In 2013, he joined the Multimedia Signal Processing Laboratory at the National Taiwan University of Science and Technology as a Post-Doctoral Fellow. In 2015, he was involved in research with the Digital Video and Multimedia (DVMM) Laboratory, Columbia University, New York. He has worked on foreground segmentation, biometrics, digital halftoning, watermarking, image compression, and enhancement. His general interests lie in machine learning and multimedia processing, and their related applications.

Dr. Liu was a recipient of the Doctoral Dissertation Excellence Awards from the Taiwanese Association for Consumer Electronics (TACE), the Institute of Information & Computing Machinery (IICM), and Image Processing and Pattern Recognition Society of Taiwan (IPPR), in 2013 and 2014, the Excellent Paper Award from the *Computer Vision, Graphics and Image Processing* (CVGIP) in 2013, and the *International Computer Symposium* (ICS) in 2014, the Master's Thesis Awards from the Taiwan Fuzzy Systems Association (TFSA) and ChiMei Optoelectronics (CMO) in 2009.

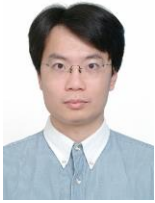

**Jing-Ming Guo** (M'04–SM'10) received the Ph.D. degree from the Institute of Communication Engineering, National Taiwan University, Taipei, Taiwan, in 2004. He is currently a Professor with the Department of Electrical Engineering, National Taiwan University of Science and Technology, Taipei, Taiwan. His research interests include multimedia signal processing, biometrics, computer vision, and digital halftoning.

Dr. Guo is a senior member of the IEEE and a Fellow of the IET. He has been promoted as a Distinguished Professor in 2012 for his significant research contributions. He received the Best Paper Award from the International Computer Symposium in 2014, the Outstanding youth Electrical Engineer Award from Chinese Institute of Electrical Engineering in 2011, the Outstanding young Investigator Award from the Institute of System Engineering in 2011, the Best Paper Award from the IEEE International Conference on System Science and Engineering in 2011, the Excellence Teaching Award in 2009, the Research Excellence Award in 2008, the Acer Dragon Thesis Award in 2005, the Outstanding Paper Awards from IPPR, Computer Vision and Graphic Image Processing in 2005 and 2006, and the Outstanding Faculty Award in 2002 and 2003.

Dr. Guo will be the General Chair of IEEE International Conference on Consumer Electronics in Taiwan in 2015, and was the Technical program Chair for IEEE International Symposium on Intelligent Signal Processing and Communication Systems in 2012, IEEE International Symposium on Consumer Electronics in 2013, and IEEE International Conference on Consumer Electronics in Taiwan in 2014. He has served as a Best Paper Selection Committee member of the IEEE Transactions on Multimedia. He has been invited as a lecturer for the IEEE Signal Processing Society summer school on Signal and Information Processing in 2012 and 2013. He has been elected as the Chair of the IEEE Taipei Section GOLD group in 2012. He has served as a Guest Co-Editor of two special issues for Journal of the Chinese Institute of Engineers and Journal of Applied Science and Engineering. He serves on the Editorial Board of the Journal of Engineering, The Scientific World Journal, International Journal of Advanced Engineering Applications, Detection, and Open Journal of Information Security and Applications. Currently, he is Associate Editor of the IEEE Transactions on Image Processing, IEEE Transactions on Multimedia, IEEE Signal Processing Letters, the Information Sciences, and the Signal Processing.